\documentclass[aps,prl,twocolumn,preprintnumbers]{revtex4-2}
\usepackage{amsmath, mathrsfs, amssymb,amsfonts,amsthm,graphicx, epsf, dcolumn, yfonts, subfigure}
\usepackage[hyperfootnotes=true]{hyperref}
\usepackage{color}
\usepackage{slashed}
\usepackage{setspace}
\usepackage{cancel}
\usepackage{wasysym}
\usepackage{float}
\pdfoutput=1
 \newcommand{\be}{\begin{equation}}
 \newcommand{\ee}{\end{equation}}
 \newcommand{\bea}{\begin{eqnarray}}
 \newcommand{\eea}{\end{eqnarray}}

\newcommand{\beq}{\begin{equation}}
\newcommand{\eeq}{\end{equation}}

\renewcommand*{\thefootnote}{\fnsymbol{footnote}}

\begin{document}

\preprint{IFT-UAM/CSIC-23-126}

\title{Reentrant phase transitions of quantum black holes}
\author{Antonia M. Frassino,$^{1,2}$ Juan F. Pedraza,$^{3}$ Andrew Svesko$^{4,5}$ and Manus R. Visser$^{6}$}
\affiliation{$^1$Departamento de F\'{i}sica y Matem\'{a}ticas, University of Alcal\'{a}, Campus universitario 28805, Alcal\'a de Henares (Madrid), Spain\\
$^2$Departament de Física Quàntica i Astrofísica and
  Institut de Ciències del Cosmos, Universitat de Barcelona, 08028 Barcelona, Spain\\
 $^3$Instituto de F\'isica Te\'orica UAM/CSIC
Calle Nicol\'as Cabrera 13-15, Cantoblanco, 28049 Madrid, Spain\\
$^4$Department of Physics and Astronomy, University College London, London WC1E 6BT, UK\\
$^5$Department of Mathematics, King’s College London, Strand, London, WC2R 2LS, UK\\
$^6$Department of Applied Mathematics and Theoretical Physics, University of Cambridge, Cambridge CB3 0WA, UK}

\begin{abstract}\vspace{-2mm}
\noindent  We show backreaction of quantum fields on black hole geometries can trigger new thermal phase transitions. Specifically, we study the phase behavior of the three-dimensional quantum-corrected static BTZ  black hole, an exact solution to specific semi-classical gravitational equations due to quantum conformal matter, discovered through braneworld holography. Focusing on the canonical ensemble, for large backreaction, we find novel reentrant phase transitions as the temperature monotonically increases, namely, from thermal anti-de Sitter space to the   black hole and back to thermal anti-de Sitter. The former phase transition is first-order,  a quantum analog of the classical Hawking-Page phase transition, while the latter   is zeroth-order and has no classical counterpart.

\end{abstract}

\renewcommand*{\thefootnote}{\arabic{footnote}}
\setcounter{footnote}{0}

\maketitle

\noindent \textbf{Introduction.} Black hole thermodynamics offers a window into the nature of quantum gravity. 
A paradigmatic example is the Hawking-Page (HP) phase transition of black holes in asymptotically anti-de Sitter (AdS) space 
\cite{Hawking:1982dh}: below a certain temperature, large AdS black holes in equilibrium with radiation give way to thermal AdS, signaling  an exchange of the dominant contribution to the quantum gravitational partition function. Originally discovered for four-dimensional Schwarzschild-AdS black holes, the HP transition  also exists for their three-dimensional counterparts
\cite{Maldacena:1998bw,Mano:1999xs,Birmingham:2002ph,Kurita:2004yn}, i.e., Ba\~{n}ados-Teitelboim-Zanelli (BTZ) 
black holes \cite{Banados:1992wn,Banados:1992gq}. 
To wit, a static BTZ black hole of mass $M$ has metric
\beq
\begin{split}
&ds^2=-f(r)dt^2+\frac{dr^2}{f(r)}+r^2d\phi^2\,,\\
&f(r)=\frac{r^2}{\ell_3^2}-8G_{3}M\,,
\end{split}
\label{btzmetric}\eeq
with $\text{AdS}_{3}$ length scale $\ell_{3}$, three-dimensional Newton's constant $G_3$, and has horizon radius $r_+^2 = 8 G_{3} M \ell_3^2.$  Via the canonical partition function, the BTZ free energy is
\begin{equation}
    F_{\text{BTZ}} =M - TS= - \frac{\pi^2 \ell_3^2}{2 G_{3}} T^2\;,
\end{equation}
for temperature $T\!=\!r_{+}/2\pi\ell^{2}_{3}$ and entropy $S\!=\!2\pi r_{+}/4G_{3}$.
Comparing to the free energy of thermal AdS, $ F_{\text{AdS}}=M_{\text{AdS}}= -1/8G_{3}$, a first-order phase transition  occurs at a temperature $T_{\text{HP}} = 1/(2 \pi \ell_3)$. 
When $T<T_{\text{HP}}$, thermal AdS has a lower free energy than the black hole, while for $T>T_{\text{HP}}$ the black hole becomes the dominant contribution to the partition function.

The study of such gravitational phase transitions has expanded to a plethora of black hole backgrounds, revealing rich physical phenomena. For instance, Reissner-Nordstr\"{o}m AdS black holes undergo a first-order phase transition between large and small black holes analogous to the liquid/gas phase change of van der Waals fluids \cite{Chamblin:1999tk,Chamblin:1999hg,Caldarelli:1999xj}, displaying the same critical behavior \cite{Kubiznak:2012wp}. Moreover, reentrant phase transitions -- a sequence of two or more phase transitions due to a monotonic change to any thermal quantity where the initial and final states are macroscopically similar -- occur for, \emph{e.g.}, $d=4$ Born-Infeld AdS black holes \cite{Gunasekaran:2012dq}, rotating AdS black holes in $d\geq6$ dimensions  \cite{Altamirano:2013ane,Ahmed:2023dnh}, or $U(1)$ charged Lovelock black holes \cite{Frassino:2014pha}, sharing traits akin to multicomponent fluids, binary gases, and liquid crystals \cite{Narayankumar}. 

Each of these studies consider classical black hole backgrounds.  
It is natural to wonder how quantum matter 
influences black hole phase transitions through semi-classical backreaction. 
A complete treatment of this question, however,  requires solving the semi-classical Einstein equations, $G_{ab} =8 \pi G \left \langle T_{ab} \right \rangle$, a difficult and open problem in higher than two spacetime dimensions.

Here we use braneworld holography~\cite{deHaro:2000wj} to exactly study phase transitions of semi-classical black holes to all orders of backreaction due to a large number of quantum fields.
In this framework a $d$-dimensional end-of-the-world brane is coupled to Einstein's general relativity in an asymptotically $(d+1)$-dimensional AdS background, which has a dual holographic interpretation as a conformal field theory (CFT) living on the AdS boundary.
In effect, the brane, typically located a small distance away from the AdS boundary, renders the (on-shell) bulk action finite 
by integrating out bulk degrees of freedom up to the brane, as in holographic regularization \cite{deHaro:2000vlm}.
This procedure induces a specific higher curvature gravity theory on the brane, coupled to a CFT with an ultraviolet (UV) cutoff that backreacts on the dynamical brane geometry. Thence,  
classical solutions to the bulk Einstein equations 
exactly correspond to solutions of the  semi-classical field equations on the brane.
Specifically, classical AdS black holes map to quantum-corrected black holes on the brane, to 
all orders of backreaction \cite{Emparan:2002px}.

 \noindent \textbf{Quantum BTZ black hole.} We will study the phase transitions of a specific braneworld model, the three-dimensional quantum BTZ (qBTZ) family of black holes~\cite{Emparan:2020znc}.  This solution follows from introducing an $\text{AdS}_{3}$   brane \cite{Karch:2000ct} inside a  static, asymptotically   AdS$_{4}$ geometry described by the C-metric \cite{Emparan:1999wa,Emparan:1999fd,Emparan:2020znc}. The brane intersects the $\text{AdS}_{4}$ black hole horizon such that the horizon localizes on the brane. 
Via braneworld holography, the backreacted geometry and horizon thermodynamics of the qBTZ are known analytically, allowing for an exact description of its phase structure. 

The metric of the quantum BTZ black hole is \cite{Emparan:2020znc} 
\beq
\begin{split}
&ds^2=-f(r)dt^2+\frac{dr^2}{f(r)}+r^2d\phi^2\,,\\
&f(r)=\frac{r^2}{\ell_3^2}-8\mathcal{G}_3M-\frac{\ell \mathcal{F}(M)}{r}\,,
\end{split}
\label{eq:qBTZ}\eeq
with horizon radius $r_{+}$ being the largest root of $f(r_{+})=0$.  Here $M$ is the mass, 
$\ell$
represents  an infrared bulk cutoff length, and 
  $\mathcal{G}_3=G_{3}/\sqrt{1+(\ell/\ell_{3})^{2}}$ is the `renormalized' Newton's constant. Note for $\ell=0$ the classical BTZ metric~\eqref{btzmetric} is recovered.
 The form function $\mathcal{F}(M)$ is found by solving the brane equations of motion \cite{Emparan:2020znc} 
 \beq
\mathcal{F}(M) =8\frac{1-\kappa x_{1}^{2}}{(3-\kappa x_{1}^{2})^{3}}\;,
 \eeq 
where $\kappa=\pm1,0$ corresponds to different brane slicings ($\kappa=-1$ gives a BTZ black hole) and $x_{1}$ is a parameter controlling the mass, see (\ref{eq:qbtzthermo}). 
Together, $(x_{1},\kappa)$ parametrize a family of brane black holes and black strings covering a finite range of masses \cite{Emparan:1999wa,Emparan:1999fd}. Classically, these solutions exist in disconnected branches of allowed masses, while quantum effects unify these branches. Lastly, the $\text{AdS}_{3}$ radius $\ell_{3}$ is related to the (induced) brane cosmological constant 
\beq \label{lambda} \Lambda_3 \equiv - \frac{1}{L_3^2} 
= -  2 \left ( \frac{1}{\ell^2} + \frac{1}{\ell_3^2} - \frac{1}{\ell} \sqrt{\frac{1}{\ell^2} + \frac{1}{\ell_3^2} } \right)\,.
\eeq

The metric (\ref{eq:qBTZ}) can be understood as a `quantum' black hole in the sense it is guaranteed to be a solution to the full semi-classical theory on the brane at all orders in quantum backreaction. The parameter $\ell$ controls     the strength of the backreaction due to the $\text{CFT}_{3}$, and  
also features in the central charge $c_{3}$ of the cutoff $\text{CFT}_{3}$   \cite{Emparan:2020znc}
\beq \label{centralcharge} c_{3} 
=\frac{\ell}{2G_{3}\sqrt{1+(\ell/\ell_3)^{2}}}\,.\eeq
For small backreaction, $\ell/\ell_{3}\ll1$, then $L_{3}^{2}\approx \ell_{3}^{2}$ while $2c_{3}G_{3} \approx\ell$. Thus, for fixed $c_{3}$, gravity becomes weak on the brane as $\ell\to0$ such that there is no backreaction due to the CFT and no renormalization of  Newton's constant. Lastly, note $\ell\approx 2c_{3}L_{\text{P}}$, where $L_{\text{P}}=G_{3}$ is the three-dimensional Planck length (with $\hbar=1$). Hence, the quantum correction in the qBTZ metric   is not a Planckian effect since $c_3 \gg 1$.

\noindent \textbf{Thermodynamics of the quantum BTZ black hole.}
The thermodynamics of the quantum BTZ black hole on the brane is inherited from the  black hole thermodynamics in the bulk. The   AdS$_4$ C-metric  describes  an accelerating black hole, however, there is no acceleration horizon in our setup since we work in the regime of `small acceleration'~\cite{Appels:2016uha}.
 Hence, there is only a black hole horizon in thermal equilibrium with its surrounding.  
The mass $M$, temperature $T$ and entropy $S$ of the classical bulk black hole are \cite{Emparan:1999fd,Emparan:2020znc} (see also~\cite{Kudoh:2004ub})
\bea
M&=&\frac{\sqrt{1+\nu^{2}}}{2G_3}\frac{z^2(1-\nu z^3)(1+\nu z)}{(1+3z^2+2\nu z^3)^2}\,, \label{eq:qbtzthermo}\\
T&=&\frac{1}{2\pi\ell_3}\frac{z(2+3\nu z+\nu z^3)}{1+3z^2+2\nu z^3}\,, \label{eq:temp}\\
S&=&\frac{\pi \ell_3\sqrt{1+\nu^{2}}}{G_3}\frac{z}{1+3z^2+2\nu z^3}\,,
\label{eq:genentropy}
\eea
where $z\!\equiv \!\ell_3/(r_{+}x_1)$ and $\nu \! \equiv \!\ell / \ell_3$ both have range $[0,\infty)$.
Each quantity may be derived by identifying the bulk on-shell Euclidean action with the canonical free energy~\cite{Kudoh:2004ub}. Previous work \cite{Appels:2016uha,Anabalon:2018ydc} examined accelerating black hole thermodynamics but not in the presence of a brane.

From the brane perspective, the qBTZ black hole has the same temperature $T$, while the four-dimensional  Bekenstein-Hawking  entropy $S$ is identified with the three-dimensional generalized entropy,  $S\equiv S_{\text{gen}}$ \cite{Emparan:2006ni,Emparan:2020znc}, accounting for both higher-curvature corrections and semi-classical matter effects. Thus, if $\ell$ and $\ell_3$ are kept fixed, the qBTZ first law takes the standard form
\beq dM=TdS_{\text{gen}}\,,\label{eq:qbtzfirstlaw}\eeq
valid to all orders in backreaction, and where the qBTZ mass is identified as $M$. Classical entropy being replaced by the generalized entropy in the first law also occurs for two-dimensional semi-classical black holes \cite{Pedraza:2021cvx,Svesko:2022txo}.

The thermal quantities   obey the Smarr relation \cite{Frassino:2022zaz}
\be \label{eq:3dsmarr}
0=TS_{\text{gen}}-2P_3V_3+\mu_3 c_3\,,
\ee
where $P_3=-\Lambda_3/(8\pi G_{3})$ is the pressure with conjugate `thermodynamic volume' $V_{3}$, and $\mu_{3}$ is the chemical potential conjugate to  $c_{3}$  
(see \cite{Frassino:2022zaz} for exact expressions in terms of $\ell_3$, $z$ and $\nu$).
 Unlike higher-dimensional Smarr formulae, the mass is absent from the three-dimensional Smarr relation (\ref{eq:3dsmarr}) since $G_3 M$ has vanishing scaling dimension, as with the classical BTZ black hole \cite{Frassino:2015oca, Frassino:2019fgr}.

 The appearance of extra thermodynamic variables in the Smarr relation (\ref{eq:3dsmarr}) suggests an extended black hole thermodynamics \cite{Kastor:2009wy}. Specifically, the first law (\ref{eq:qbtzfirstlaw}) generalizes to include pressure and central charge variations  
\beq dM=TdS_{\text{gen}}+V_{3}dP_{3}+\mu_{3}dc_{3}\;.\label{eq:extfirstlaw}\eeq
In the context of braneworld holography, extended thermodynamics of black holes on the brane is naturally induced from the standard thermodynamics of bulk black holes including work done by the brane, \emph{e.g.,} dynamical pressure $P_{3}$ corresponds to variable brane tension \cite{Frassino:2022zaz}. Here, however, we focus on the canonical ensemble, defined by fixing $(T,P_{3},c_{3})$. 

 \begin{figure}[t!]
\centering
\includegraphics[width=0.465\textwidth]{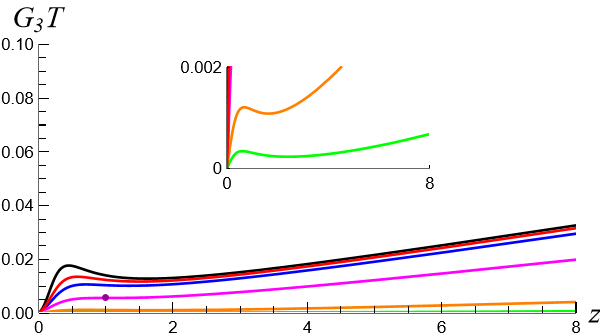}
\caption{\label{fig:tempvsz} Temperature of qBTZ black hole at $c_3\!\! \!=\!\!\! 10$ and various pressures (from bottom to top): $G_{3}^{3}P_{3}\!=\!\{5.0\times10^{-7}\,(\text{green}),5.0\times10^{-6}\,(\text{orange}),G_3^3P_{\text{crit}}\!\approx \! 5.83\times 10^{-5}\,(\text{magenta}),1.33\times 10^{-4} \,(\text{blue}), 1.61\times10^{-4}\,(\text{red}),1.81\times 10^{-4} \,(\text{black})\}$. 
The critical line inflection point is in purple.}
\end{figure}

 In FIG. \ref{fig:tempvsz} we plot temperature as a function of $z$ at fixed $P_{3}$ and $c_3$. For all pressure -- except a critical one  $P_3=P_{\text{crit}}$ --  the temperature  has two turning points, signifying three branches: (A) `cold' black hole, $z\in(0,z_{\text{max}})$, with $z_{\text{max}}$ denoting the local maximum of $T$; (B) `intermediate' black hole, $z\in( z_{\text{max}},z_{\text{min}})$, with $z_{\text{min}}$ marking the local minimum of $T$, and (C) `hot' black hole, $z\in(z_{\text{min}},\infty)$.  (This terminology is motivated by the end behavior of $T(z)$, but, note branch B black holes can have temperatures less than those in branch A.)
For $\nu>1$, $z_{\text{max}}=\nu^{-1/3}$  and $z_{\text{min}}$ is the positive root of  $\nu z^{3}+3z^{2}-3\nu z-1=0$, and conversely, for $\nu<1$, $z_{\text{min}}=\nu^{-1/3}$ while $z_{\text{max}}$ is the positive root of the same equation; for any $\nu$, $z_{\text{max}}<z_{\text{min}}$, except when $\nu=1$, where $z_{\text{max}}=z_{\text{min}}=z_{\text{crit}}$ (see below). Notably, for $\nu>1$, branch A typifies black holes with masses $0<M<1/24 \mathcal G_3$, while B and C  branches have $-1/8 \mathcal G_3< M <0$ (with $M=0$ at $z=0,z_{\text{max}}$). Alternately, for $\nu<1$, branches A and B have $0<M<1/24 \mathcal G_3$, and branch C has $-1/8 \mathcal G_3< M <0$ (with $M=0$ at $z=0,z_{\text{min}}$) \footnote{Our three   branches do not coincide with branches $1a$, $1b$, and $2$ characterizing the mass in \cite{Emparan:2020znc}. Precisely, let $z_{1}$ be the positive root to $4 z^3 \nu + 3   z^2 -1=0$, $z_{2}$ be the positive root of $\nu z^{3}+3z^{2}-3\nu z-1=0$, and $z_{3}=\nu^{-1/3}$ (these coincide with \cite{Kudoh:2004ub} upon $z\to \nu z$ and $\lambda=\nu^{2}$). The value $z_{1}$ denotes where the mass  has a maximum $M=1/24 \mathcal G_3$, and $M=0$ at $z_3$. The heat capacity \eqref{eq:heatcapcano} diverges at $z_2$ and $z_3$, and is zero at $z_1$. For any $\nu$ there are two branches of positive mass black holes with $\kappa=-1$: branch $2$, where $0<z<z_{1}$, or $\sqrt{3}<x_{1}<\infty$, and branch $1b$, where $z_{1}<z<z_{3}$, or $0<x_{1}<\sqrt{3}$. For $\nu \neq 0$ there is also one branch of negative mass black holes with $\kappa=+1$: branch $1a$, where $z_{3}<z<\infty$, or $0<x_{1}<1$. We can compare to our branches by identifying  $z_2 = z_{\text{min}}$ and $z_3 = z_{\text{max}}$ for $\nu>1$, and $z_2 = z_{\text{max}}$ and $z_3 = z_{\text{min}}$ for $\nu < 1$. Thus, our three branches characterizing $T(z)$ are distinguished by $ z_{\text{min}}$ and  $z_{\text{max}}$, whereas the three branches of $M(z)$ in \cite{Emparan:2020znc} are separated by $z_1$ and $z_3$.}.
In contrast, classical BTZ has a single branch of black holes since $T$ is monotonic in $r_+$.

Further, for qBTZ the  temperature  has an inflection point when both the first and second $z$-derivatives of $T$ at fixed $\nu$ vanish. 
This occurs when $z_{\text{crit}}=\nu_{\text{crit}}=1$ and yields
the following critical pressure and temperature 
\beq P_{\text{crit}}=\frac{1}{16\pi c_{3}^{2}G_{3}^{3}(2+\sqrt{2})}\;,\quad T_{\text{crit}}=\frac{1}{4\sqrt{2}\pi c_3 G_{3}}\;.\label{eq:critPT}\eeq
Both expressions depend on the fixed central charge. Alternately, we can fix the pressure, yielding the critical central charge given by the inverse of \eqref{eq:critPT}, $c_{\text{crit}} \propto 1/\sqrt{P_3}$.

\noindent \textbf{Phase transitions of the quantum BTZ black hole.}
We focus on thermal phase behavior of the quantum BTZ black hole in 
the canonical ensemble with free energy 
 \beq
 \begin{split}
 \label{freeenergy}
F_{\text{qBTZ}}&\equiv M-TS_{\text{gen}}\\
&=-\frac{z^{2}\sqrt{1+\nu^{2}}}{2G_{3}}\frac{[1+2\nu z+\nu z^{3}(2+\nu z)]}{(1+3z^{2}+2\nu z^{3})^{2}}\;.
\end{split}
\eeq
Here we expressed  the free energy in terms of $\nu$,  but we can use \eqref{lambda} and \eqref{centralcharge} to rewrite $\nu$ in terms of $P_3$ and $c_3$
\beq
\label{eq:nu}
\nu =\frac{\sqrt{32 \pi c_3^2 G_3^3 P_3 (1- 8 \pi c_3^2 G_3^3 P_3)}}{1- 16 \pi c^2_3 G_3^3 P_3}\,.
\eeq
Observe $\nu$ diverges when $c_3^2 G_3^2 P_3 = 1/16 \pi$, placing an upper bound on pressure, $P_3 < 1/16\pi c_3^2 G_3^3$.
Also note the canonical ensemble is equivalent to a fixed $\nu$ ensemble.

  \begin{figure*}[!t]
\centering
\includegraphics[width=12cm,height=6cm]{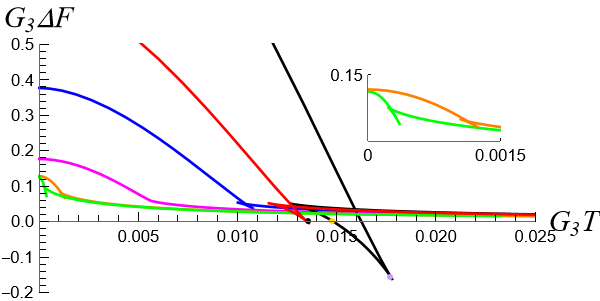}
\caption{\label{fig:free}
Free energy difference of the qBTZ black hole and quantum thermal AdS at   $c_{3}=10$ and various pressures $P_3$ (same values as in FIG. \ref{fig:tempvsz}, ordered left to right). For $P_3\neq P_{\text{crit}}$ the free energy as a function of the temperature has inverse swallow-tail behavior and consists of three branches. 
For small pressures, thermal AdS (with $\Delta F =0$) always dominates. For   $G_{3}^{3}P_3 > 1.6135\times10^{-4}$ (red, center curve), thermal AdS  dominates for small   and large temperatures, and in between    the intermediate branch has the lowest free energy. A first-order phase transition between thermal AdS and the intermediate branch occurs if this branch intersects the $\Delta F =0$ line (yellow point), and a zeroth-order phase transition  occurs at the right cusp (purple point). The phase transitions coincide when the right cusp  intersects $\Delta F=0$ (black point).
}
\end{figure*}

Analogous to the classical case \cite{Hawking:1982dh}, the qBTZ black hole can transition into thermal AdS, since at fixed pressure and central charge the black hole can evaporate due to Hawking radiation, and, conversely, thermal AdS can transition into a qBTZ black hole via collapse of the thermal gas. Thus, to analyze the phase behavior, we compare the free energy of the qBTZ black hole to that of `quantum' thermal AdS$_3$ (qTAdS) geometry, i.e., 
 pure $\text{AdS}_{3}$ including backreaction due to the cut-off $\text{CFT}_{3}$ on the brane. This geometry contains thermal radiation in equilibrium at an arbitrary temperature since the CFT$_3$ is taken to be in a thermal state.  Explicitly, the  qTAdS  geometry takes the form~(\ref{eq:qBTZ}), with $\mathcal{F}(M)=0$ and $M_{\text{qTAdS}}=-1/8\mathcal{G}_{3}$, coinciding with the $z\to\infty$ limit of the qBTZ solution. Notice that while backreaction does not alter the form of the metric from classical $\text{AdS}_{3}$, the $\text{CFT}_{3}$ makes itself felt through the parameter $\nu$ in $\mathcal G_{3}$. Thus,  the sole effect of backreaction due to conformal matter in thermal $\text{AdS}_{3}$ is to renormalize Newton's constant. An analogous result occurs for semi-classical Jackiw-Teitelboim gravity: backreaction due to conformal matter does not break the symmetries of the $\text{AdS}_{2}$ geometry; only the dilaton (or, equivalently, the effective two-dimensional Newton constant) receives quantum corrections (e.g., \cite{Pedraza:2021cvx}).
 Further, in this limit the generalized entropy \eqref{eq:genentropy} vanishes, a consequence of quantum fluctuations renormalizing $G_{3}$ (capturing the spirit of \cite{Susskind:1994sm}), and the free energy \eqref{freeenergy} becomes $F_{\text{qTAdS}}=M_{\text{qTAdS}} =-1/8\mathcal{G}_{3}$.

 In FIG. \ref{fig:free} we make a parametric plot of the free energy difference  $\Delta F \!\equiv \! F_{\text{qBTZ}}-F_{\text{qTAdS}}$ versus temperature $T$  (using $z$ as the parameter).
At $P=P_{\text{crit}}$ the free energy plot is smooth.
 However, for all positive values $P\neq P_{\text{crit}}$ the  free energy  diagram shows   inverse swallow-tail behavior and contains three different branches: cold, intermediate and hot black hole branches, corresponding to the branches in FIG. \ref{fig:tempvsz}.
 The cold branch begins at small temperature and ends at the lower-right cusp ($z=z_{\text{max}}$), whereas the hot branch corresponds to the   `horizontal' line that extends from  the upper-left cusp ($z=z_{\text{min}}$) off to high temperature. The intermediate branch is the curve connecting these two cusps. 
Since the free energy $F_{\text{qBTZ}}$ and temperature go to zero as $z \rightarrow 0$, the difference $\Delta F$ does not vanish at zero temperature; instead it starts from a positive value that depends on $P_3$ and $c_3$. 
 
 The free energy plot displays   phase transitions between thermal AdS and the qBTZ black hole for a certain pressure and temperature range.  
  Since we are subtracting the free energy of thermal AdS, the black hole branches below the horizontal axis of the plot in FIG.~\ref{fig:free} are the only ones that are thermodynamically favored with respect to thermal AdS.  
  Everywhere else, thermal AdS has a lower free energy than the black hole.  When the right cusp  intersects   $\Delta F=0$, that is the starting point of the phase transitions. The temperature and pressure where this occurs are found by solving when $\Delta F=0$ and $z=z_{\text{max}}$,  
  at $\nu=3\sqrt{3}$ and $z=1/\sqrt{3}$, giving
  \begin{align}
   &T_{0}=\frac{\sqrt{3}}{2\pi\ell_{3}}=\frac{9}{8\sqrt{7}\pi c_{3}G_{3}}\;,\label{eq:HPtemp}\;\quad P_{0}=\frac{14-\sqrt{7}}{224\pi c_{3}^{2}G_{3}^{3}}\;.
   \end{align}
   Note $T_{0}$ is larger than the classical HP temperature. 
  
\begin{figure}[t!]
\begin{centering}
\hspace{-9mm}\includegraphics[width=8.5cm]{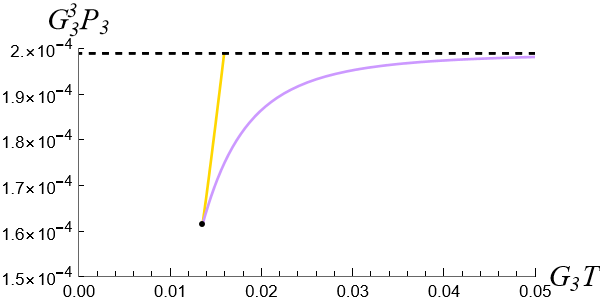}
\caption{$P_{3}$ versus $T$ phase diagram at fixed $c_{3}=10$. The dashed black line corresponds  to $\nu\to\infty$. 
The straight yellow and curved purple lines denote lines of first- and zeroth-order phase transitions, respectively. In the region between the yellow and purple curves the qBTZ black hole is thermally favored; thermal AdS dominates the canonical ensemble outside this region. The   (black) cusp where the yellow and purple lines meet corresponds to $G_{3}^{3}P_{0}=1.6135\times10^{-4}$ and $G_3 T_0 = 0.0135$.
 The isolated critical point lies at $G_{3}^{3}P_{\text{crit}}\!\approx \!5.83\times10^{-5}$ and $G_{3}T_{\text{crit}} \approx 5.63\times10^{-3}$ (not shown). 
}
\label{fig:PvT}
\end{centering}
\end{figure}

  For larger pressures, as the temperature monotonically increases, there are \emph{reentrant phase transitions} from thermal AdS to qBTZ and back to thermal AdS. The former transition occurs when branch B intersects the $\Delta F =0$ line. Since there is a discontinuity in the slope of the free energy, this is a first-order phase transition,  a quantum analog of the Hawking-Page phase transition. The latter transition between the branch B of qBTZ and thermal AdS occurs at the right cusp. Since there is a jump discontinuity in the free energy, this is a zeroth-order phase transition. Thus, the   reentrant phase transition is   described by the combination of the first- and zeroth-order phase transitions as the temperature monotonically varies. In the $P_{3}$ vs. $T$ phase diagram at fixed $c_3$ (FIG.~\ref{fig:PvT})  we depict  coexistence lines of first- and zeroth-order phase transitions. The (black) intersection point of the two phase transitions \eqref{eq:HPtemp} is neither representative of a second-order phase transition or a critical point.

 Notably, such reentrant phase transitions do not occur for the classical BTZ black hole.
 As noted, here the zeroth-order phase transitions only occur for large enough $P_{3}$, or, correspondingly, $\nu > 3 \sqrt{3}$, i.e., large backreaction.  In this regime, the brane has decreasing tension and the gravitational theory on the brane becomes more massive and effectively four-dimensional \cite{Emparan:2020znc}.

\begin{figure}[t!]
\begin{centering}
\hspace{-9mm}\includegraphics[width=8.5cm]{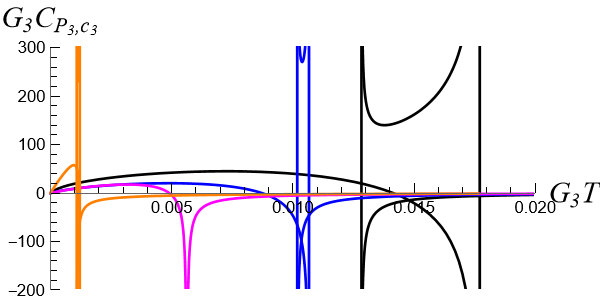}
\caption{  Heat capacity at fixed pressure and central charge for select parameters in FIG. \ref{fig:tempvsz} (from left to right: $5.0\times10^{-6}\,(\text{orange}),G_3^3P_{\text{crit}}\!\approx \! 5.83\times 10^{-5}\,(\text{magenta}),1.33\times 10^{-4} \,(\text{blue}),1.81\times 10^{-4} \,(\text{black})$).
}
\label{fig:heatcapacity}
\end{centering}
\end{figure}

The heat capacity allows us to determine which branches are stable under thermal fluctuations \cite{Kudoh:2004ub} 
\begin{align} &C_{P_{3},c_{3}} = T\left(\frac{\partial S_{\text{gen}}}{\partial T}\right)_{\hspace{-1mm} P_{3},c_{3}} \label{eq:heatcapcano} \\
&=\frac{\sqrt{1+\nu^{2}}\pi\ell_{3}}{2G_{3}(1-\nu z^{3})}\frac{z(2+3\nu z+\nu z^{3})(3z^{2}-1+4\nu z^{3})}{(3z^{2}-1-3\nu z+\nu z^{3})(1+3z^{2}+2\nu z^{3})}\;. \nonumber \end{align}
In FIG. \ref{fig:heatcapacity} we plot the heat capacity (\ref{eq:heatcapcano}) versus temperature at fixed $P_3$ and $c_3$. For $T\neq  T_{\text{crit}}$  the cold black hole branch has partly positive and negative heat capacity, while the intermediate branch has $C_{P_{3},c_{3}} >0$ and the hot branch has $C_{P_{3},c_{3}}<0$.
The heat capacity vanishes for a $z_1$ that is the positive root of $4 z^3 \nu + 3   z^2 -1=0$ (for $\nu \neq 1$), and diverges at $z_\text{min}$ and $z_{\text{max}}$, where $z_1 < z_{\text{max}} < z_{\text{min}}.$
Note that the quantum Hawking-Page transition is between qTAdS and a `stable' black hole branch, as in the classical HP transition.

\noindent \textbf{Discussion.} We used braneworld holography to study the thermal phase structure of black holes corrected due to semi-classical backreaction. Aside from a quantum counterpart of the first-order Hawking-Page transition, the qBTZ black hole also undergoes zeroth order phase transitions -- a feature solely due to semi-classical effects. The physical viability of the zeroth-order phase transition is questionable, since such transitions typically do not occur in nature, for thermodynamically stable systems. A zeroth-order phase transition may be indicating we are missing a novel phase with lower free energy. For instance, the black hole could transition into a system with additional degrees of freedom or ``hair'' that are not captured by the qBTZ solution. We leave this for future investigation. 

The free energy of the braneworld black holes investigated here was previously derived in \cite{Kudoh:2004ub} from the four-dimensional bulk perspective, however, that analysis lacked a physical interpretation in terms of 
quantum black holes and reentrant phase transitions were not observed. Our analysis on the brane carries over to the bulk and thus suggests   this is the first example of  a   semiclassical black hole    undergoing such a thermal reentrant phase transition.  Further, we emphasize that the type of reentrant phase transition we uncover differs from those which have appeared in previous literature
\cite{Gunasekaran:2012dq,Altamirano:2013ane,Ahmed:2023dnh,Frassino:2014pha}, which describe transitions between different phases of the black hole, e.g., from large to small and back to large black holes. Meanwhile, the reentrant phase
transition we find is from thermal AdS to the black hole and back to thermal AdS, i.e., a reentrant Hawking-Page phase transition. Notably, only for large backreaction do the reentrant phase transitions occur. This suggests such features are unlikely to be found via standard perturbative techniques in studying quantum backreaction \cite{Steif:1993zv,Casals:2016ioo,Casals:2019jfo}.

Since the brane geometry is asymptotically AdS, the brane gravity theory has a holographic interpretation in terms of a two-dimensional (defect) CFT, where all $1/c$  corrections are accounted for in a large central charge-$c$ expansion. Thus, via this second layer of AdS/CFT duality, the phase transitions of the qBTZ black hole should have a dual interpretation \cite{Frassino:2022zaz}. Classically, the HP phase transition is argued to be dual to the (de)confinement transition of a large-$c$ conformal gauge theory \cite{Witten:1998zw}. 
Our analysis implies the phase structure of the thermal $\text{CFT}_{2}$ dual to the qBTZ black hole drastically changes when including all $1/c$ effects.
Namely, in the canonical ensemble the temperature of the first-order phase transition changes and a new zeroth-order phase transition arises, leading to reentrant transitions between (de)confined phases. It would be worth studying these new features from 
a microscopic perspective \cite{KordZangeneh:2017lgs}. The bulk and brane system, moreover, is dual to a boundary $\text{CFT}_{3}$ ($\text{BCFT}_{3}$) \cite{Takayanagi:2011zk,Fujita:2011fp}. Since the qBTZ free energy is equal to the bulk black hole free energy, thus having the same phase behavior,
the dual $\text{BCFT}_{3}$ should exhibit reentrant phase transitions. It would be interesting to study this further.

Our study offers many future explorations. Firstly, we focused on the  canonical ensemble. 
In fact, the static qBTZ black hole has four different ensembles (at fixed temperature) to examine. A study of ensembles at fixed thermodynamic volume (initiated in \cite{Johnson:2023dtf}) will shed new light on the instability of `superentropic' black holes \cite{Johnson:2019mdp, Hennigar:2014cfa}, its microscopic interpretation \cite{Johnson:2019wcq}, along with other inequalities constraining (extended) thermodynamic variables \cite{Cvetic:2010jb,Amo:2023ixe}.
Further, our analysis can be generalized to other quantum black holes, \emph{e.g.}, rotating and charged qBTZ \cite{Emparan:2020znc,Climent:2024nuj}, or quantum de Sitter black holes \cite{Emparan:2022ijy,Panella:2023lsi}. Adding rotation or charge will enrich the phase structure. 
Lastly, we focused on black holes in three-dimensions. 
It is natural to wonder how to generalize to higher dimensions. So far finding higher-dimensional braneworld black holes has proven challenging (cf. \cite{Tanahashi:2011xx}). 
Perhaps progress can be made using a large-dimension limit of (bulk) general relativity \cite{Emparan:2013moa,Emparan:2020inr}, to construct higher-dimensional static braneworld black holes, as done for evaporating black holes \cite{Emparan:2023dxm}.

\noindent \emph{Acknowledgements.}
We are grateful to Roberto Emparan, Robie Hennigar, Clifford Johnson, Jorge Rocha and Marija Toma\v{s}evi\'{c} for useful discussions and correspondence.  AMF is supported by AGAUR grant 2017-SGR 754, and State Research Agency of
MICINN through the “Unit of Excellence Mar\'ia de Maeztu 2020-2023” award to the Institute of Cosmos Sciences (CEX2019-000918-M). JFP is supported by the `Atracci\'on de Talento' program grant 2020-T1/TIC-20495 and by the Spanish Research Agency through the grants CEX2020-001007-S and PID2021-123017NB-I00, funded by MCIN/AEI/10.13039/501100011033 and by ERDF A way of making Europe. AS is supported by STFC grant ST/X000753/1 and was partially supported by the Simons Foundation via \emph{It from Qubit Collaboration} and EPSRC as this work was being completed. MRV is supported by  SNF Postdoc Mobility grant P500PT-206877 ``Semi-classical thermodynamics of black holes and the information paradox''.

\bibliographystyle{apsrev4-2}
\bibliography{qptrefs}

\end{document}